\documentclass[prl,aps,twocolumn,superscriptaddress]{revtex4-2}
\usepackage{graphicx,color}
\usepackage{amsthm}
\usepackage{amsfonts}
\usepackage{algorithmic}
\usepackage{enumerate}
\usepackage{latexsym}
\usepackage{amsmath}
\usepackage{amssymb}
\usepackage{subfig}
\usepackage[colorlinks=true,citecolor=blue,linkcolor=blue]{hyperref}

\usepackage{dcolumn}% Align table columns on decimal point
\usepackage{bm}% bold math
\usepackage[justification=raggedright]{caption}
\usepackage[T1]{fontenc}
\usepackage{mathptmx}

\begin{document}

\title{Pairing correlation of the Kagome-lattice Hubbard model
with the nearest-neighbor interaction}

\author{Chen Yang}
\affiliation{School of Physics and Astronomy, Beijing Normal University, Beijing 100875, China\\}
\author{Chao Chen}
\affiliation{Department of Basic Courses, Naval University of Engineering, Wuhan 430033, China\\}
\affiliation{School of Physics and Astronomy, Beijing Normal University, Beijing 100875, China\\}
\author{Runyu Ma}
\affiliation{School of Physics and Astronomy, Beijing Normal University, Beijing 100875, China\\}
\author{Ying Liang}
\affiliation{School of Physics and Astronomy, Beijing Normal University, Beijing 100875, China\\}
\author{Tianxing Ma}
\email{txma@bnu.edu.cn}
\affiliation{School of Physics and Astronomy, Beijing Normal University, Beijing 100875, China\\}
\affiliation{Key Laboratory of Multiscale Spin Physics(Ministry of Education), Beijing Normal University, Beijing 100875, China\\}

\begin{abstract} 
A recently discovered family of Kagome lattice materials, $\emph{A}\mathrm{V}_{3}\mathrm{Sb}_{5}$($\emph{A}$= $\mathrm{K,Rb,Cs}$),
has attracted great interest, especially in the debate over its dominant superconducting pairing symmetry.
To explore this issue, we study the superconducting pairing behavior within the Kagome-Hubbard model through the constrained path Monte Carlo method.
It is found that doping around the Dirac point generates a dominant next-nearest-neighbour-$d$ pairing symmetry driven by on-site Coulomb interaction $U$.
However, when considering the nearest-neighbor interaction $V$, it may induce nearest-neighbor-$p$ pairing to become the preferred pairing symmetry.
Our results provide useful information to identify the dominant superconducting pairing symmetry in $\emph{A}\mathrm{V}_{3}\mathrm{Sb}_{5}$ family.
\end{abstract}

\maketitle

%\section{Introduction}
\noindent
\underline{\it Introduction}~~
The discovery of superconducting $\emph{A}\mathrm{V}_{3}\mathrm{Sb}_{5}$ family has attracted  tremendous attention due to their exotic properties\cite{Yu2021,sciadv.abl4108,Xu_2024,Li_2024,YuanWang,ChinPhysLett.39.047402}. These materials offer an important platform to explore a variety of novel quantum phases of electron correlations\cite{PhysRevLett.125.247002,mine2024directobservationanisotropiccooper,PhysRevLett.127.046401,Mielke2022,Nie2022,PhysRevB.105.174518}, and their physical properties are closely related to their unique electronic structure. Specifically, the crystal structure of $\emph{A}\mathrm{V}_{3}\mathrm{Sb}_{5}$ forms a layered Kagome system. Accompanying the strong frustration in the Kagome lattice, its electronic structure has flat bands and Dirac cones, which give rise to a number of intriguing properties, including quantum anomalous Hall effect\cite{PhysRevB.105.235134,PhysRevB.85.144402,Mi2023}, charge density wave order\cite{Xu2022,PhysRevX.13.031030}, $Z_{2}$ topological invariant\cite{PhysRevX.11.031026,PhysRevLett.127.046401}, and superconductivity \cite{Yin2019,profe2024kagomehubbardmodelfunctional}. Among them, the possible dominant superconducting pairing symmetry is highly debate. Thermal conductivity measurements and a V-shaped pairing gap suggest an unconventional nodal superconductivity in $\emph{A}\mathrm{V}_{3}\mathrm{Sb}_{5}$\cite{zhao2021nodalsuperconductivitysuperconductingdomes,Kang2022,FENG20211384,PhysRevLett.127.217601}. On the other hand, measurements of magnetic penetration depth and the effects of magnetic impurities suggest the absence of sign-change in the superconducting order parameter\cite{PhysRevLett.127.187004,Duan2021,PhysRevResearch.4.023215}, pointing to a nodeless $s$-wave superconductivity. Therefore, reconciling the contradictions in experimental evidence was a significant challenge to overcome.

To study the superconductivity in $\emph{A}\mathrm{V}_{3}\mathrm{Sb}_{5}$ family, a Hubbard-type model defined on a Kagome lattice has been studied intensively\cite{PhysRevB.104.165127,PhysRevLett.95.037001,PhysRevB.105.L100502,PhysRevB.105.245131,PhysRevB.104.075148,PhysRevB.89.125129}. For example, $s$-wave superconductivity, driven by on-site Coulomb interactions, has been proposed in the Kagome-Hubbard model by using the determinant quantum Monte Carlo method\cite{PhysRevResearch.5.023037}. In correlated electron systems, the van Hove singularities (VHSs) in the density of states may play an essential role in novel superconducting phenomena \cite{PhysRevB.90.245114}. In a Kagome lattice, calculations around upper van Hove filling by random-phase-approximation indicate that $d$-wave or $f$-wave pairing symmetries are likely more favorable\cite{PhysRevLett.127.177001,PhysRevLett.110.126405}.
Additional research through an analytic renormalization group analysis has suggested $d +id$ superconductivity in the Kagome-lattice Hubbard model around the van Hove filling,
with enhancements attributed to long-range Coulomb interactions, which reduce sublattice interference effects\cite{PhysRevB.86.121105,PhysRevB.87.115135,PhysRevB.109.075130}.
Furthermore, it has been theoretically proposed that this chiral $d +id$ superconductivity can even become the dominant order in the 1/6 hole-doped Kagome lattice by using the variational cluster approach\cite{PhysRevB.85.144402}. These theoretical discoveries intertwined with experimental results have deepened our understanding of pairing states and mechanisms.
However, different methods lead to different results and conflicting experimental evidence has caused ongoing controversy. Due to the extensive experimental controversy and 
theoretical proposal regarding the superconducting pairing and its mechanism in $\emph{A}\mathrm{V}_{3}\mathrm{Sb}_{5}$ compounds, identify the pairing symmetry of the superconducting order parameter by some exact numerical method is urgently required\cite{PhysRevB.86.121105,PhysRevB.109.054504,Wilson_2024,10.1093/nsr/nwac199,PhysRevB.106.174514,Neupert2022}.

In this paper, we employ the constrained path Monte Carlo (CPMC) method to study the pairing correlation in the Kagome-lattice Hubbard model. Simulations are mainly performed on $3 \times 6^{2}$ lattices with periodic boundary conditions. We present our unbiased numerical results of ground-state pairing correlation functions for different pairing symmetries. Our results show that the on-site Coulomb interaction $U$ causes the next-nearest neighbor(NNN) $d$-wave(NNN-$d$) pairing to dominate over other pairing symmetries. However, as we introduce the nearest-neighbor(NN) interaction with $V=-1.0$, the system favors the nearest-neighbor $p$-wave(NN-$p$) pairing. The rest of this paper is organized as follows. Section 2 presents the model we are going to investigate, along with the computational methodology employed. In section 3, the pairing correlations are discussed. A summary is made in section 4.

\noindent
\underline{Theoretical method}~~
The Kagome-lattice Hubbard model is described by the Hamiltonian
\begin{eqnarray}
H=-t\sum_{<i,j>\sigma }^{}c^{\dagger } _{i\sigma } c_{j\sigma }+U\sum_{i}n_{i\uparrow }n_{i\downarrow  } +V\sum_{<i,j>\sigma }^{}n_{i\sigma } n_{j\sigma }.
\end{eqnarray}
Here, $c^{\dagger }_{i\sigma }$ ($ c_{i\sigma }$) creates (annihilates) electrons at site $i$ in the Kagome
lattice with spin $\sigma\left ( \sigma=\uparrow,\downarrow \right )$. $t$ represents the nearest-neighbor hopping integral. $ n_{i\sigma }=c^{\dagger }_{i\sigma }c_{i\sigma }$ denotes the electron number of spin $\sigma$ at site $i$. $\left \langle i,j \right \rangle$ denotes the NNs. $U$ is the on-site Hubbard interaction, while the interaction between NN sites is denoted by $V$. In our computational process, we set the hopping amplitude $t=1$.

\begin{figure*}
\centering
\includegraphics[width=0.8\textwidth]{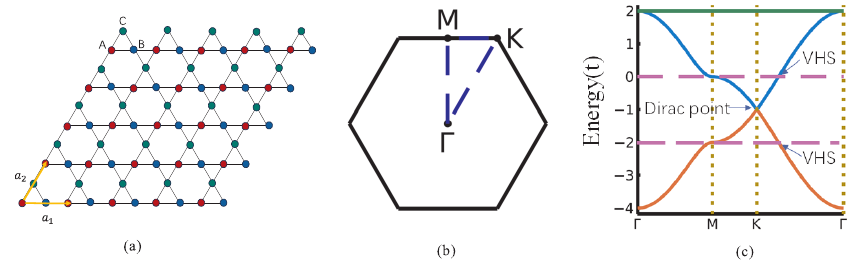}
\caption{(a)The Structure of the Kagome lattice with lattice vectors $a_{1}$ and $a_{2}$. In each unit cell, the sublattice is labeled as A(red dots), B(blue dots) and C(green dots). (b) The Brillouin zone of the Kagome lattice. (c) The band structure of the tight-binding Kagome model.}
    \label{Fig1}
\end{figure*}

As observed, each site is surrounded by four nearest-neighbor sites, four next-nearest neighbor sites, and six third-nearest neighbor (3rd NN) sites in the Kagome lattice. Fig.~\ref{Fig1}(a) shows the structure of the Kagome lattice. Fig.~\ref{Fig1}(b) displays the Brillouin zone of the Kagome lattice, where $K$, $M$, and $\Gamma$ represent the high-symmetry points. As seen in Fig.~\ref{Fig1}(c), the Kagome lattice features three bands, including two cone-shaped Dirac bands and one distinctive flat band. At $K$ point, there is a Dirac point. While at $M$ point, there are two van Hove singularities.

\begin{figure}
\centering
\includegraphics[width=0.45\textwidth]{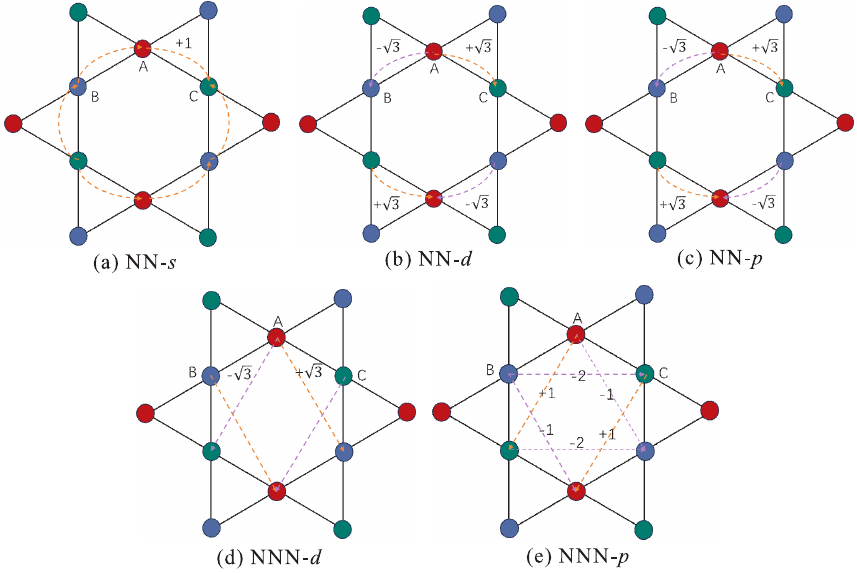}
\caption{Pairing symmetries between the sites in the Kagome lattice: (a)NN-$s$, (b)NN-$d$, (c)NN-$p$, (d)NNN-$d$, (e)NNN-$p$. The orange and purple arrows represent positive and negative pairing, respectively.}
    \label{Fig2}
\end{figure}

To study the superconducting pairing behavior, we define the pairing correlation function as follows:
\begin{eqnarray}
C_{\alpha } \left ( r \right )=\frac{1}{N_{cor}\left ( r \right )  } \times \sum_{ij}^{}\delta \left ( \left | R_{i}-R_{j}\right |-r  \right ) \left \langle \Delta^{\dagger }  _{\alpha }\left ( i \right ) \Delta _{\alpha }\left ( j \right )   \right \rangle,
\end{eqnarray}
where $N_{cor}\left ( r \right )=\sum_{ij}^{}\delta \left ( \left | R_{i}-R_{j}\right |-r  \right )$ stands for the number of correlations whose length is $r$, and $\alpha$ represents different pairing symmetry. The corresponding order parameter for spin-triplet and spin-singlet pairings  $\Delta^{\dagger }  _{\alpha }\left ( i \right ) $ is defined as
\begin{eqnarray}
\Delta^{\dagger }_{\alpha }\left ( i \right )=\sum_{l}f^{\dagger }_{\alpha}\left ( \delta _{l}  \right )\left ( c_{i\uparrow }c_{i+\delta _{l} \downarrow }\pm c_{i\downarrow  }c_{i+\delta _{l} \uparrow }  \right )^{\dagger }.
\end{eqnarray}
Here, $f_{\alpha}\left ( \delta _{l}  \right )$ is the form factor of pairing function and the vectors $\delta _{l}$ denotes the connections between sites. We referenced five kinds of pairing forms from the Kagome lattice\cite{PhysRevLett.127.177001}, which are plotted in Fig.~\ref{Fig2}. The pairing function in momentum space was converted into real-space pairing using the Fourier transform and lattice harmonics. This figure displays the real-space pairing symmetries between the sites.
 
To identify the dominant pairing symmetry in the investigated system, we utilize the CPMC method to study the long-range part of the pairing correlation function in the Kagome-Hubbard model, which can effectively extract ground state properties while avoiding the sign problem \cite{PhysRevB.89.125129,PhysRevLett.78.4486,PhysRevB.55.7464,PhysRevLett.127.177001,Liu_2018,Fang_2021}. It extracts the ground state from an initial guess by exploring different paths in the Slater determinant space\cite{PhysRevB.89.125129}. To address potential sign issues, we employ a constrained-path approximation in the CPMC algorithm \cite{PhysRevLett.127.177001,Liu_2018}. In the
CPMC method, extensive benchmark calculations showed that the systematic error induced by the constraint is within a few percent and the ground-state observables are insensitive to the choice of trial wave function. In
our CPMC simulations, we employ closed-shell electron fillings and use the corresponding free-electron $U = 0$ wave function as the trial wave function\cite{PhysRevB.106.134513,PhysRevB.107.245106,PhysRevB.84.121410}. In a typical run, we set the average number of random walkers to be 1200 and the time step $\Delta \tau  = 0.05$. We perform measurements in 40 blocks of 320 Monte Carlo steps to ensure statistical independence.

\noindent
\underline{\it Results \& Discussions}~~
\begin{figure*}
\centering
\includegraphics[width=0.9\textwidth]{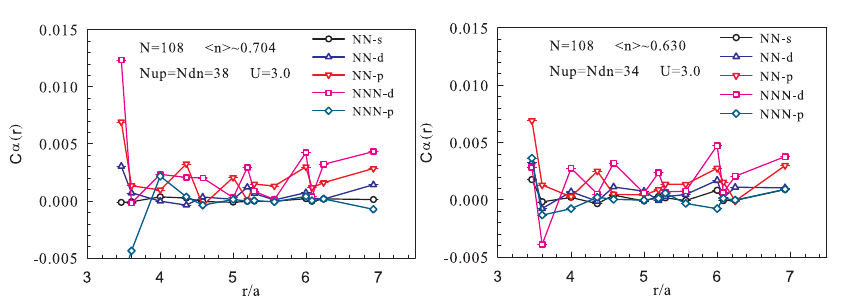}
\caption{Pairing correlations $C_{\alpha } $ as a function of distance $r$ for different pairing symmetries with $U = 3.0$ at (a) $\left \langle n \right \rangle \approx 0.704$ (left) and (b) $\left \langle n \right \rangle \approx 0.630$ (right) and $L = 6$.}
    \label{Fig3}
\end{figure*}
We first examine the dominant pairing symmetry for a fixed $U=3.0$ at different electron fillings. In Fig.~\ref{Fig3}, we present the long-range part of the pairing correlations $C_{\alpha }$ with different symmetries as a function of distance $r$ on the $3 \times 6^{2}$ lattices at the electron filling $\left \langle n \right \rangle \approx0.704$ and $\left \langle n \right \rangle \approx0.630$ corresponding to closed-shell around the Dirac point $\left \langle n \right \rangle \approx 0.667$. In Fig.~\ref{Fig3}(a), the electron filling is set to $\left \langle n \right \rangle \approx 0.704$, and in Fig.~\ref{Fig3}(b) is set to $\left \langle n \right \rangle \approx 0.630$. Both Fig.~\ref{Fig3}(a) and Fig.~\ref{Fig3}(b) display that the NNN-$d$ pairing symmetry plays a dominant role with on-site interaction $U=3.0$. As can be seen in Fig.~\ref{Fig3}(a), the NN-$s$ oscillates around zero at long-range distances, indicating that the NN-$s$ pairing correlation between the nearest neighbor electrons is very weak or nearly absent. Our findings suggest that doping around the Dirac point generates a dominant NNN-$d$ pairing symmetry driven by on-site Coulomb interaction $U$. In our present work, we are mainly concerned with the pairing correlation function versus distance for different pairing symmetries. For all long-range distances between electron pairs, the competition between $d$-wave and $p$-wave correlation plays a crucial role in the selection of the dominant pairing symmetry. By thoroughly considering the effects of both NN and NNN pairings, we demonstrate that the NNN-$d$ pairing symmetry dominates, which provide a guideline for understanding the superconducting pairing symmetries in the Kagome-Hubbard model. 

\begin{figure}
\centering
\includegraphics[width=0.45\textwidth]{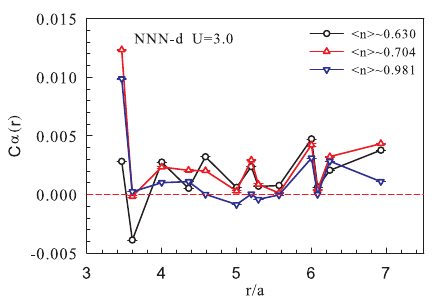}
\caption{Long-range part of NNN-$d$  pairing correlations with different electron fillings at $\left \langle n \right \rangle \approx0.630$, $\left \langle n \right \rangle \approx0.704$, and $\left \langle n \right \rangle\approx0.981$  at $U = 3.0$.}
\label{Fig4}
\end{figure}

To further explore the long-range part of the dominant NNN-$d$ pairing correlation function versus distance, we set varying electron filling values to $\left \langle n \right \rangle \approx 0.630$, $0.704$, and $0.981$. In Fig.~\ref{Fig4}, we present $C_{\alpha } \left ( r \right )$ as a function of distance $r$ for different electron fillings, with NNN-$d$ pairing correlations shown at $U= 3.0$. Figure \ref{Fig4} demonstrates that the superconducting pairing with NNN-$d$ symmetry is the weakest with hole doping at $\left \langle n \right \rangle \approx 0.981$, which indicates that the NNN-$d$ pairing symmetry is suppressed when the system is doped in proximity to half-filling. The NNN-$d$ pairing correlation begins to drop quickly from $r\approx 3.4$ and reaches its minimum value at $r\approx 3.6$ for different electron fillings. Additionally, we observe that the value of NNN-$d$ pairing correlation function is consistently positive at $\left \langle n \right \rangle \approx 0.704$.

\begin{figure}
\centering
\includegraphics[width=0.45\textwidth]{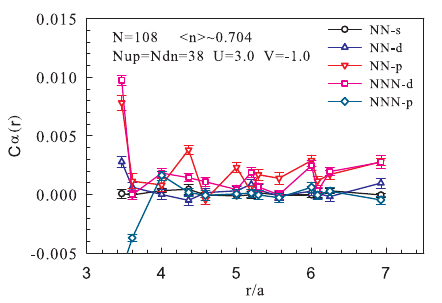}
\caption{Pairing correlations $C_{\alpha }\left ( r \right )$ as a function of distance $r$ for different pairing symmetries with $U = 3.0$ and $V= -1.0$.
}
\label{Fig5}
\end{figure}

To study the effect of the NN interaction $V$  on pairing symmetries, we calculated the
$C_{\alpha }\left ( r \right )$ for different pairing symmetries at $\left \langle n \right \rangle \approx 0.704$, $U = 3.0$ and $V = -1.0$ in Fig.~\ref{Fig5}. As the data illustrated, applying $V$ enhances the NN-$p$ pairing. This indicates that the NN interaction $V$ causes the NN-$p$ surpasses the NNN-$d$ and take the lead in pairing symmetry. Based on the above analysis, we speculate that the negative $V$ affects the NN-$p$ symmetry and the NNN-$d$  symmetry differently, leading to a shift in the dominant pairing symmetry.

\begin{figure*}
\centering
\includegraphics[width=0.95\textwidth]{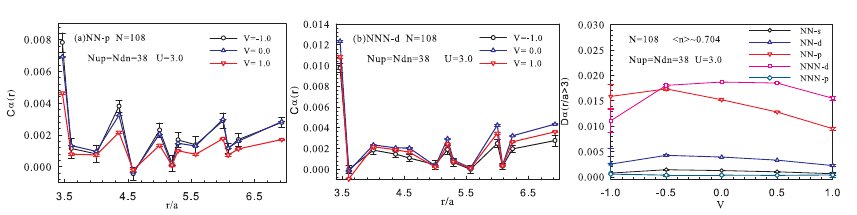}
\caption{Pairing correlations $C_{\alpha}\left ( r \right )$ as a function of distance $r$ for (a) NN-$p$ (b)NNN-$d$ with different $V$ at $U$ = 3.0. (c)Pairing correlations $D_{\alpha}(r/a > 3)$ as a function of the nearest-neighbor interaction $V$ at $U$ = 3.0 on the L = 6 lattice for different pairing symmetries.}
\label{Fig6}
\end{figure*}

As illustrated in Fig.~\ref{Fig6}(a), comparing the results for the NN interaction $V$=-1.0, 0.0, and 1.0, it is evident that the strength of the NN-$p$ pairing symmetry is enhanced with decreasing $V$. Further observing Fig.~\ref{Fig6}(b), we find that when $V=0.0$, the NNN-$d$ pairing correlation function is the largest. Figure~\ref{Fig6}(c) presents the long-range parts of various pairing correlations as a function of $V$, in which we sum up the correlations whose distance is three times larger than lattice constant $a$, and we denote it by $D_{\alpha}(r/a > 3)$. It can be concluded that a negative $V$ enhances the NN-$p$ pairing, while both positive and negative $V$ suppress the NNN-$d$ pairing. This elucidates why, at $V=-1.0$, the NN-$p$ symmetry supersedes the NNN-$d$ symmetry, emerging as the dominant one.

\begin{figure}
\centering
\includegraphics[width=0.45\textwidth]{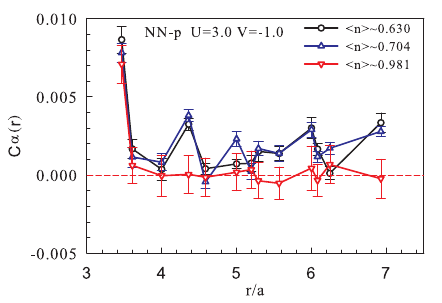}
\caption{Long-range part of NN-$p$ pairing correlations with different electron fillings at $\left \langle n \right \rangle \approx 0.630$, $\left \langle n \right \rangle \approx 0.704$, and $\left \langle n \right \rangle \approx 0.981$ at $U=3.0$ and $V=-1.0$.}
\label{Fig7}
\end{figure}

Figure~\ref{Fig4} displays the long-range part of NNN-$d$ pairing correlations with different electron fillings at $U=3.0$ and $V=0.0$. Applying $V=-1.0$ makes the NN-$p$ pairing symmetry dominate. Therefore, we focus on the analysis of the pairing correlations of NN-$p$ symmetry at different electron fillings with $U=3.0$ and $V=-1.0$. Figure~\ref{Fig7} illustrates that the NN-$p$ pairing correlation is the weakest with hole doping close to half-filling but reaches its maximum at $\left \langle n \right \rangle \approx 0.704$. The trends in the variation of $C_{\alpha }\left ( r \right )$  at $\left \langle n \right \rangle \approx 0.630$ and $\left \langle n \right \rangle \approx 0.704$ show a considerable degree of similarity. Furthermore, at $\left \langle n \right \rangle \approx 0.981$, the NN-$p$ pairing correlation oscillates around zero for distances $r$ greater than approximately 3.5.

\begin{figure}
\centering
\includegraphics[width=0.45\textwidth]{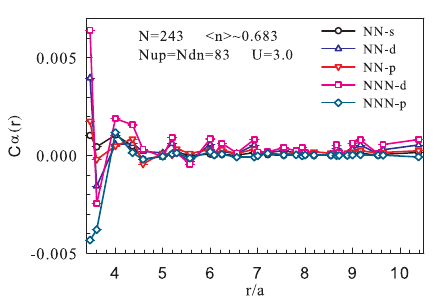}
\caption{Pairing correlations $C_{\alpha }\left ( r \right )$ as a function of distance $r$ for different pairing symmetries with $U = 3.0$ at $N_{up}=N_{dn}=83$ and $L = 9$.}
\label{Fig8}
\end{figure}

Finally, we study the effects of system size on pairing. Our simulations are performed on a larger lattice of $L = 9$ at $U=3.0$ and $\left \langle n \right \rangle \approx 0.683$, with a closed-shell filling of $N_{up}=N_{dn}=83$. The results are shown in Fig.~\ref{Fig8}, with the impact of onsite repulsive interaction $U$, the NNN-$d$ pairing symmetry dominates is consistent with our previous results. This further confirms that the finite-size effect does not affect the dominant pairing symmetry.

%\section{Conclusion}
\noindent
\underline{\it Conclusions}~~
 In summary, we have studied the superconducting pairing correlation of the Kagome-lattice Hubbard model using the CPMC method. In calculating the pairing correlation functions, we were mainly concerned with the closed-shell case. Our numerical results show that in the presence of on-site Coulomb interaction $U$, the NNN-$d$ pairing symmetry plays a dominant role. Upon introducing the NN interaction $V$, NN-$p$ pairing may surpass NNN-$d$ pairing and become the dominant pairing symmetry. Moreover, our result indicates a very weak lattice-size effect. When considering the nearest-neighbor interaction $V$, our findings suggest a novel competition between NN-$p$ and NNN-$d$ pairing correlations. In all, our results provide a deeper understanding of superconducting pairing mechanisms in the Kagome lattice model.

\noindent
\underline{\it Acknowledgements:}
This work was supported by Beijing Natural Science
Foundation (No. 1242022).  The numerical simulations in this work were performed at HSCC of Beijing Normal University. 
\bibliography{ref}

\end{document}